\documentclass[reprint,twocolumn,aps,prb,floats,superscriptaddress]{revtex4-1}
\usepackage[english]{babel}
\usepackage[ansinew]{inputenc}
\usepackage{amsmath}
\usepackage{amssymb}
\usepackage{color}
\usepackage{graphicx}

\definecolor{darkgreen}{RGB}{0,100,50}

\begin{document}
\preprint{APS/123-QED}
\title{Adsorption Sites of Individual Metal Atoms on Ultrathin MgO(100) Films}
\author{Edgar Fernandes}
\author{Fabio Donati}
\author{Fran\c{c}ois Patthey}
\affiliation{Institute of Physics, Ecole Polytechnique F{\'e}d{\'e}rale de Lausanne (EPFL), Station 3, CH-1015 Lausanne, Switzerland}
\date{\today}
\author{Srdjan Stavri\'c}
\affiliation{Vin\v{c}a Institute of Nuclear Sciences (020), University of Belgrade P.O.Box 522, 11001 Belgrade, Serbia}
\author{\v{Z}eljko \v{S}ljvan\v{c}anin}
\affiliation{Vin\v{c}a Institute of Nuclear Sciences (020), University of Belgrade P.O.Box 522, 11001 Belgrade, Serbia}
\affiliation{Texas A\&M University at Qatar, Doha, Qatar}
\author{Harald Brune}
\affiliation{Institute of Physics, Ecole Polytechnique F{\'e}d{\'e}rale de Lausanne (EPFL), Station 3, CH-1015 Lausanne, Switzerland}

\begin{abstract}
We use Ca doping during growth of one and two monolayer thick MgO films on Ag(100) to identify the adsorption sites of individual adatoms with scanning tunneling microscopy. For this we combine atomic resolution images of the bare MgO layer with images of the adsorbates and the substitutional Ca atoms taken at larger tip-sample distance. For Ho atoms, the adsorption sites depend on MgO thickness. On the monolayer, they are distributed on the O and bridge sites according to the abundance of those sites, $1/3$ and $2/3$ respectively. On the MgO bilayer, Ho atoms populate almost exclusively the O site. A third species adsorbed on Mg is predicted by density functional theory and can be created by atomic manipulation. Au atoms adsorb on the bridge sites for both MgO thicknesses, while Co and Fe atoms prefer the O sites, again for both thickness.
\end{abstract}
\maketitle

\section{Introduction}
Single atoms on ultrathin insulating layers grown on metal surfaces have spectacular magnetic properties~\cite{hei04, ott08, rau14}, in particular very long spin-coherence~\cite{bau15a} and even longer spin-relaxation times~\cite{don16, pau16, nat17} lending single adatom qubits and memories feasible. Moreover, they exhibit multiple stable charge states~\cite{rep04, ols07, ste15}, when adsorbed in the vicinity of defects they may catalyse chemical reactions~\cite{wor04}. Additionally, on surfaces of bulk oxides they are astonishingly stable~\cite{nov12} and currently considered as single atom catalysts~\cite{yan13, par13, blie15}. These remarkable properties emerge from the interaction of the atom with the surface depending critically on the adsorption site. Knowing this site is therefore mandatory to understand the thermal stability, catalytic properties, charge state, and finally the symmetry of the crystal field that determines the lifetime of magnetic quantum states~\cite{hub14}.

Field ion microscopy reveals the adsorption site, but is limited to strongly bound species on metal surfaces~\cite{wan89}. Scanning tunneling microscopy (STM) and atomic force microscopy (AFM) are more versatile and now widely employed to determine the adsorption sites of adatoms. However, all examples in the literature are limited to single element surfaces~\cite{bru90, rep03, don13, eel13}. On the surfaces of ionic crystals or thin films, such as MgO or NaCl, one often ignores which of the two sublattices gives rise to the atomic STM and AFM contrast. Density functional theory (DFT) calculations report contradicting results about the STM contrast on MgO/Ag(100)~\cite{lop04, mal14, che14}, while AFM contrast of NaCl was interpreted based on molecular markers for which adsorption geometry was obtained from DFT~\cite{lam10, teo11}. Specific to STM, the tunnel parameters required for atomic resolution on insulating layers imply very small tip-sample distances. Under these conditions, adsorbed atoms are frequently displaced or even desorbed, which further complicates the determination of their adsorption site. For example, light adsorbates such as H can get displaced by tip-sample interactions even under moderate tunnel conditions yielding fictitious adsorption sites~\cite{kle03}. As an alternative approach, electron paramagnetic resonance (EPR) was used to indirectly indentify the adsorption site of Au on thick MgO layers~\cite{yul06}. However, for the same atoms on 3 ML of MgO on Ag(100), the adsorption site remains debated~\cite{ste07b, hon07, pac05}. These issues are general for any single atom on the surfaces of ionic crystals or thin films and call for a direct and reliable experimental method.

Here we introduce dilute Ca doping to mark the Mg sublattice in STM images of one and two monolayer thick MgO films grown on Ag(100). To determine the orientation and size of the atomic MgO lattice, we record atomic resolution images on adsorbate-free areas. This lattice is overlaid onto STM images of the Ca dopants and adsorbates taken at larger tip-sample distance. On this grid all Ca atoms are on identical sites demonstrating the reliability of our method to mark the Mg positions. Comparison of the adsorbates' positions with this MgO grid unequivocally identifies the adsorption sites of Ho, Au, Co, and Fe adatoms as function of MgO thickness. These four examples are motivated by Ho being the first single atom magnet~\cite{don16}, the adsorption sites of Au being debated~\cite{pac05, ste07b, hon07}, Co having the highest possible magneto-crystalline anisotropy for a 3$d$ element~\cite{rau14}, and finally Fe on MgO being the first system where electron spin resonance (ESR) with the STM was demonstrated~\cite{bau15a} and spin-coherence times measured on a single atom. DFT calculations reveal the charge transfer and binding energy of the adsorbates on mono- and bilayer MgO/Ag(100), as well as on the (100) surface of bulk MgO.

We start by giving details on the experiment and on the density functional theory (DFT) calculations. The results and discussion section is divided into five parts. Section \textbf{A} focuses on the characterization of the pristine and Ca-doped MgO thin films. Sections \textbf{B} and \textbf{C} describe the experimental determination of the adsorption sites of Ho, as well as  STM atomic manipulation experiment on these atoms. Sections \textbf{D} and \textbf{E} present results on the adsorption site of Co, Fe, and Au. Section \textbf{IV} concludes the manuscript.

\section{Technical details}
\subsection{Experimental}
The Ag(100) surface was prepared in ultra high vacuum by repeated cycles of Ar$^+$ sputtering (800~eV, $10 \; \mu{\rm A}/{\rm cm}^2$) and subsequent annealing to 770~K. MgO thin films were grown by evaporating Mg from a Knudsen cell under a partial pressure of oxygen of $1 \times 10^{-6}$~mbar and with the sample kept at 770~K, as described in Ref.~\onlinecite{pal14}. These conditions yields an MgO growth rate of about 0.1 monolayers per minute. We define one monolayer (ML) as one MgO(100) unit cell per Ag(100) substrate atom. Calcium-doped MgO films were prepared by co-evaporating Ca and Mg under the conditions described above and with a significant lower Ca than Mg flux, adjusted to obtain the desired dopant concentration. Ho, Co, Fe, and Au atoms were evaporated from {\it e}-beam evaporators onto the sample in the STM at $T_{\rm dep} \approx 10$~K and $p < 1 \times 10^{-10}$~mbar. STM measurements were performed with a home-built STM at $T_{\rm STM} = 4.7$~K using W or PtIr tips~\cite{gai92}. Differential conductance $(\text{d}I/\text{d}V)$ spectra were acquired with a \mbox{Lock-In} amplifier using a bias modulation at 1397 Hz and working at closed feedback loop to minimize the tip-surface interaction at large biases.
\subsection{Density Functional Theory calculations}
The DFT calculations for Ho adatoms on MgO(100)/Ag(100) were carried out using the Wien2k computer code~\cite{wien2k}, with the same computational setup as the one described in Ref. \onlinecite{don16}, \textit{i.e.}, using the generalized gradient approximation (GGA) and on-site Coulomb interactions. DFT calculations of the Co and Au adatoms on thin MgO(100) films on Ag(100) were performed with the GPAW code \cite{enk10}, based on the real space grid implementation of the projector augmented wave (PAW) method~\cite{blo94,mor05}. Exchange-correlation effects were described employing the Perdew-Burke-Ernzerhof functional (PBE)~\cite{per96}. For Co, the calculations were performed within the GGA+U approach \cite{ani91,dud98}, which combines the standard PBE exchange-correlation functional with on-site Coulomb interaction, using a U value of 2~eV. The MgO(100)/Ag(100) surface was modeled with a 3$\times$3 unit cell containing 9 Mg and 9 O atoms per MgO(100) layer, placed on a three-layer metal slab with 9 Ag atoms per fcc(100) layer. In all calculations we used theoretically optimized Ag lattice constants of 4.14~{\AA}, grid spacing of 0.15~{\AA}, two-dimensional periodic boundary conditions parallel to the (100) surfaces, and 16 Monkhorst-Pack {\bf k}-points for sampling of Brillouin zone~\cite{mon76}. Open boundary conditions are applied perpendicular to the surface with 7~{\AA} of vacuum separating the oxide/metal slabs from the cell boundaries. To increase numerical stability of the calculations, the electronic states were occupied according to the Fermi-Dirac distribution with a broadening of 0.1~eV. Atomic positions were relaxed using the BFGS algorithm \cite{liu89}.

\section{Results and Discussion}
\subsection{STM characterization of pristine and\\Ca-doped MgO thin films}
\begin{figure}[b]
\includegraphics{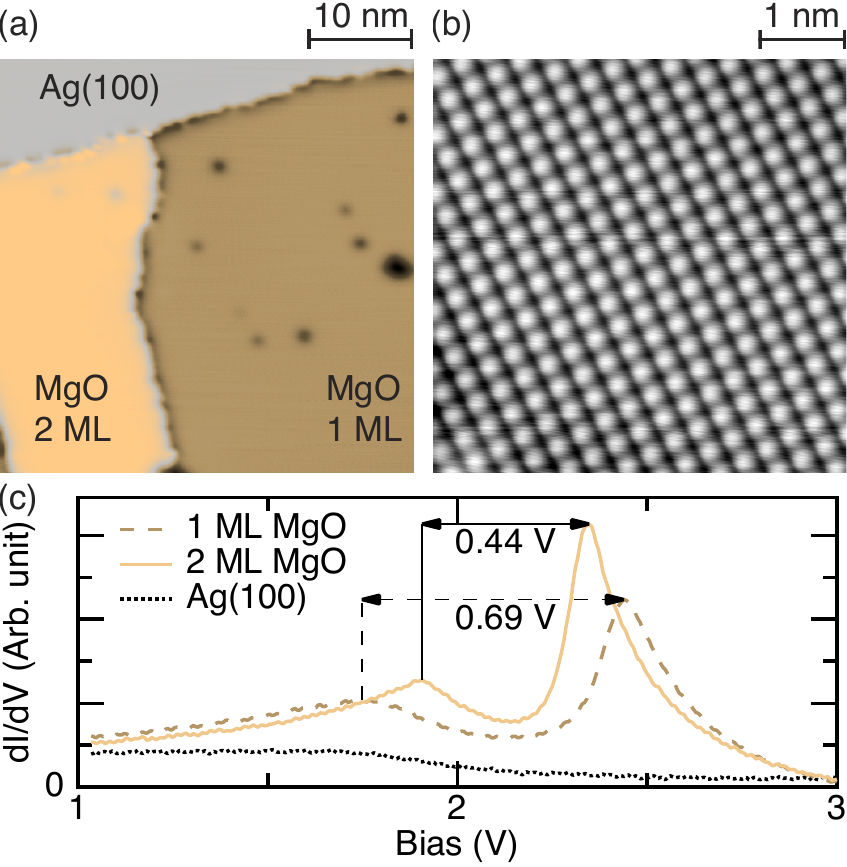}
\caption{(a) STM image of Ag(100) partially covered by MgO ($V_{\rm t} = 1$~V, $I_{\rm t} = 100$~pA). Dark spots are attributed to point defects in the oxide at the interface. (b) Atomically resolved image of 2~ML MgO ($V_{\rm t} = -10$~mV, $I_{\rm t} = 10$~nA). (c) Field emission resonance spectra recorded above 1 and 2~ML MgO as well as clean Ag(100) ($V_{\rm t} = 1$~V, $I_{\rm t} = 100$~pA, peak-to-peak modulation amplitude $V_{\rm mod} = 10$~mV).}
\label{subs}
\end{figure}

Figure~\ref{subs}(a) shows an STM image of the MgO/Ag(100) surface with the bare substrate coexisting with MgO layers of two thicknesses. Since their apparent heights are strongly bias-dependent, and sometimes inverted with respect to the expectation from morphology~\cite{bau14},  we use the energies of field emission resonances to determine the MgO thickness~\cite{sch01}. The d$I$/d$V$ spectra in Fig.~\ref{subs}(c) exhibit two resonances with distinct energy separations of 0.69~V and 0.44~V; the first is characteristic for the MgO mono-, and the second for the bilayer~\cite{sch01}. Note that a very recent paper proposes an MgO thickness calibration~\cite{pau16} that differs by one layer from the one used in the current literature and also in the present study. However, our method to determine the adsorption sites is independent of the MgO thickness.

The atomically-resolved STM image of 2~ML MgO of Fig.~\ref{subs}(b) shows a square lattice of protrusions representing one ionic sub-lattice~\cite{sch01, ste07b}. The period of $2.90 \pm 0.03$~{\AA} agrees very well with the Ag(100) nearest neighbor distance of 2.89~{\AA}. In addition, the STM image shows no superstructure, such as moir{\'e} patterns or dislocations. Both observations provide direct evidence for the MgO(100) film being uniformly and compressively strained by 3~\% to form a pseudomorphic $(1 \times 1)$ structure on Ag(100). This confirms early diffraction studies~\cite{wol99} that reveal that this lateral compression leads to a vertical expansion of the unit cell by 3.6~\%~\cite{val02}. Whether the protrusions in this image represent the Mg or the O species has been a matter of debate in theory~\cite{lop04, mal14, che14}. Our Ca doping method introduced hereafter determines it unequivocally for the respective STM tip and tunnel parameters.

\begin{figure}[tb!]
\includegraphics[width = 1.0\linewidth]{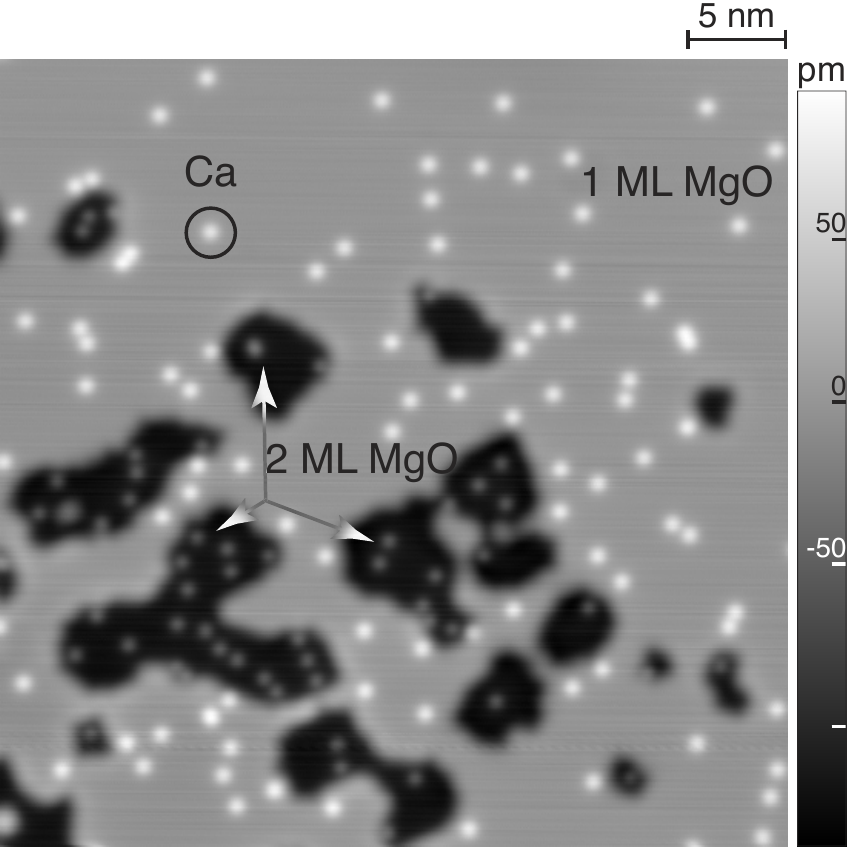}
\caption{STM image of one and two monolayers of MgO with the substitutional Ca atoms appearing as protrusions ($V_{\rm t} = -515$~mV, $I_{\rm t} = 100$~pA). Note that the apparent height of the MgO layers is inverted to their thickness.}
\label{Ca} 
\end{figure}

Figure~\ref{Ca} shows an overview image of $1$ and $2$~ML Ca-doped MgO. For the employed tunnel parameters, Ca atoms are imaged as small protrusions with apparent heights of $73 \pm 2$~pm and $61 \pm 2$~pm, respectively, on 1 and 2~ML MgO. The extremely narrow apparent height distribution found for each of the two MgO thicknesses indicates that all Ca atoms are on identical lattice sites (uncertainties are calculated from the standard deviation of the object's apparent height, \textit{i.e.}, $88$ and $50$ objects respectively on $1$ and $2$~ML MgO).
In addition, the very similar apparent heights observed on 1 and 2 ML MgO suggests that all Ca atoms are localized in the topmost MgO layer. Ca atoms buried in the second layer would appear with different apparent height and be located on the other lattice site. Therefore Ca protrusions always mark the Mg lattice positions irrespectively of the local MgO thickness. Our conclusion of Ca surface segregation is supported by the literature\cite{sta05, tas85, mcc83}. Upon annealing of Ca-rich MgO crystals, Mg atoms at the surface are replaced with Ca. For our employed Ca-doping of the order of 0.5\% pratically all the Ca atoms are sufficiently far from each other to appear as individual and identical protrusions, see Figures~\ref{Ca} and~\ref{sites}.

\subsection{Adsorption site determination of Ho adatoms on thin MgO films}
\begin{figure}[t]
\includegraphics{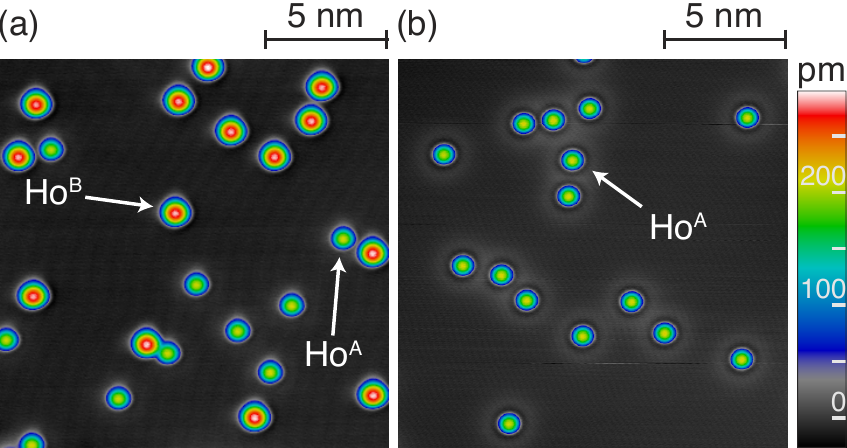}
\caption{STM images of (a) 1~ML and (b) 2~ML undoped MgO after the adsorption of $5 \times 10^{-3}$~ML of Ho. Two species, Ho$^{\rm A}$ and Ho$^{\rm B}$, are discerned by their apparent heights of $220 \pm 4$~pm {\it vs.} $295 \pm 3$~pm ($T_{\rm dep} \approx 10$~K, $V_{\rm t} = -20$~mV, $I_{\rm t} = 20$~pA).}
\label{spec}
\end{figure}

Figures~\ref{spec}(a) and (b) show $5 \times 10^{-3}$~ML of Ho deposited at 10~K on one and two monolayer of MgO/Ag(100). While two species with characteristic apparent heights coexist on 1~ML MgO, on 2~ML almost exclusively the species with smaller apparent height (Ho$^{\rm A}$) occurs.

\begin{figure}[t]
\includegraphics{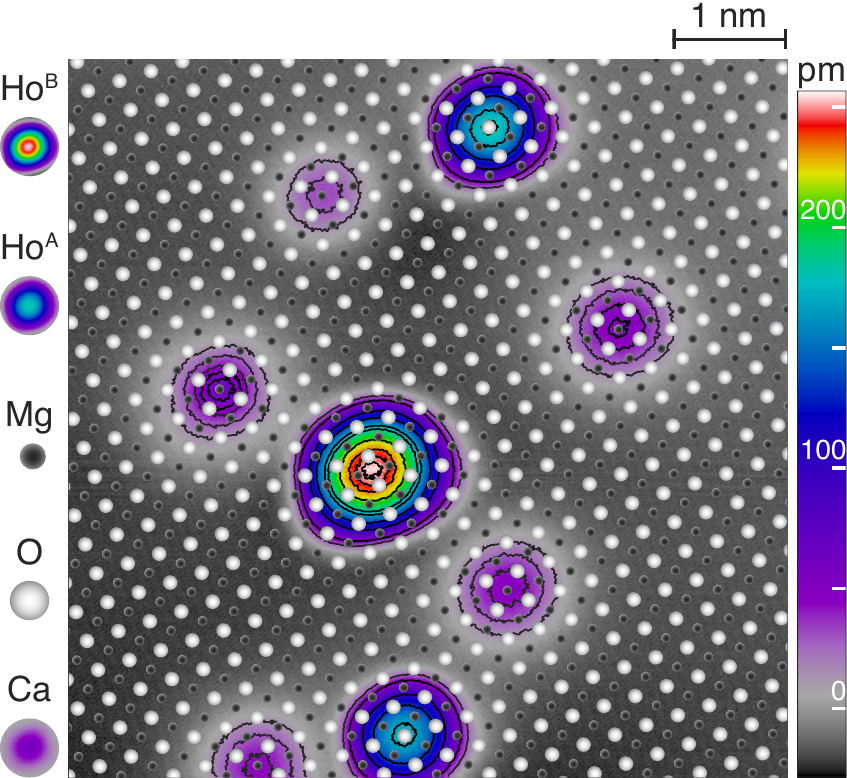}
\caption{STM image of 1 ML MgO grown with $5 \times 10^{-3}$~ML Ca doping and the same amount of Ho atoms adsorbed onto it. The orientation and spacing of the overlaid MgO lattice were determined from an atomically resolved image recorded on a bare MgO spot of the same sample. This lattice was then translated to bring one of its Mg atoms in coincidence with one of the Ca species ($V_{\rm t} = -20$~mV, $I_{\rm t} = 20$~pA).}
\label{sites}
\end{figure}

Figure~\ref{sites} shows a single MgO layer with substitutional Ca atoms, as well as both adsorbed Ho species. Calcium atoms appear as faint spots with an apparent height of $67 \pm 12$~pm for this tunneling setpoint. Note that the higher uncertainty stems from the standard deviation obtained from the 5 Ca protrusions.
The tunneling conditions yielding atomic resolution on MgO move the Ho atoms and therefore, we imaged a bare MgO spot of the same sample with atomic resolution and extracted the orientation and lattice constant of the MgO lattice from it. This lattice was then overlaid onto Fig.~\ref{sites} and one of its Mg atoms aligned with one of the substitutional Ca atoms. All other Ca species fall exactly onto Mg sites illustrating the precision of the alignment. This technique has been applied on images up to $20 \times 20$~nm$^2$ containing up to 12 Ca atoms (not shown here) with the same reliability. \color{black} Comparing the Ho positions with the overlayed MgO lattice for the shown image and for many additional ones, we infer that Ho$^{\rm A}$ adsorbs on O while Ho$^{\rm B}$ is on a bridge site. Thus, the preferred adsorption site on 2 MgO layers is on top of oxygen. In agreement, our DFT calculations identify this site for 2 ML MgO/Ag(100) as the most stable one, see Table~\ref{dft}. They also show that this site is favored for MgO(100) bulk. Therefore, from 2 ML MgO on, the ensemble of Ho atoms is dominated by a single species. This facilitates the interpretation of ensemble measurements, such as X-ray magnetic circular dichroism~\cite{don16}.

The very different abundance of both Ho species on 1 and 2~ML MgO can be traced back to MgO thickness dependent dissipation of the adsorption energy. On 1~ML MgO, the abundance of both species (Ho$^{\rm A}$: $35.8 \pm 1.6$~\% / Ho$^{\rm B}$: $64.2 \pm 1.6$~\%) reflects the one of their adsorption sites since there are two times more bridge than O sites. This is compatible with statistical growth, where the atoms come to rest at their site of impact, irrespective of its adsorption energy. The clear preference of the O site for 2~ML MgO (Ho$^{\rm A}$: $91.5 \pm 1.7$~\% / Ho$^{\rm B}$: $8.5 \pm 1.7$~\%) implies adatom motion. Thermal mobility can be ruled out since both species are immobile for hours up to 50~K. Therefore the Ho atoms exhibit transient mobility, at least from the bridge toward the adjacent O site\cite{gao12}. The fact that this occurs more readily on 2 than on 1 ML MgO is related to the dissipation of the adsorption energy via electron-hole pair excitation in the substrate~\cite{bun15, bru15} that is more efficient for atoms adsorbing on thinner MgO layers.

\begin{figure}[b!]
\includegraphics[width = 1.0\linewidth]{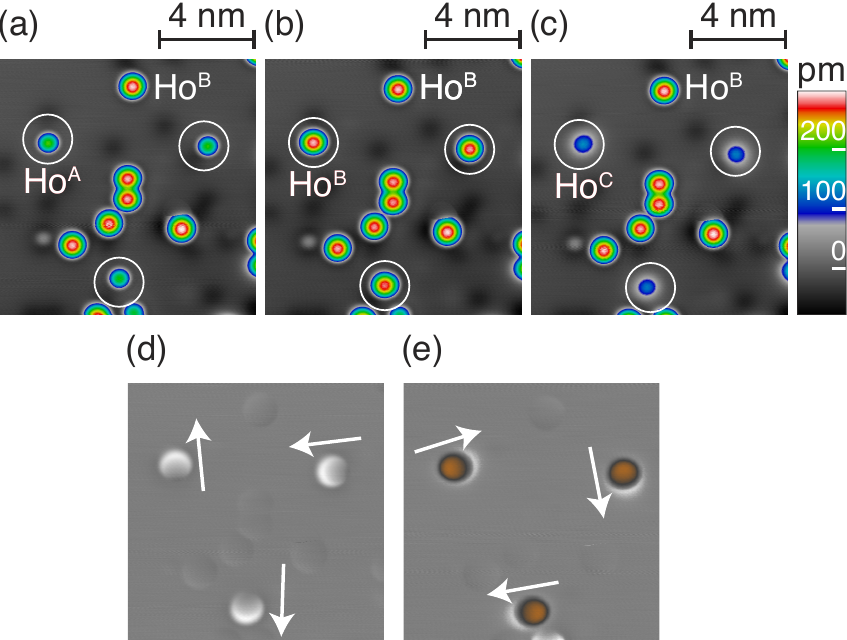}
\caption{STM images of Ho atoms on 1 ML MgO. Atoms indicated with circles are successively transformed from Ho$^{\rm A}$ (a) to Ho$^{\rm B}$ (b), and finally to Ho$^{\rm C}$ (c) by applying negative voltage ramps, see text for details (a-c: $V_{\rm t} = -100$~mV, $I_{\rm t} = 20$~pA). (d) Subtraction of images (a) from (b) using the unchanged atoms for precise alignment. (e) $\text{(b)}-\text{(c)}$. The white (brown) color indicates levels above (below) the zero plane. Arrows indicate the directions of the atomic displacements.}
\label{manip} 
\end{figure}

\subsection{STM manipulation of Ho adsorption sites}
On 1~ML MgO/Ag(100), the DFT calculations identify the Mg site as the most favorable one for Ho atoms. Although Ho atoms do not spontaneously reach this site after deposition, we can populate it by atomic manipulation. For the atomic manipulation we apply voltage ramps with the STM tip placed above the Ho atoms. To prevent major modifications of the probed area by high electric fields, the voltage is ramped while keeping the feedback loop closed, {\it i.e.}, the tunneling current stays constant while the tip retracts smoothly. Abrupt changes in the vertical position of the tip detected during the ramp evinces a modification or a displacement of the probed atom~\cite{rep04}. Figures~\ref{manip}(a-c) illustrate the result of this process on a few selected Ho atoms on 1 ML MgO. Ramping the bias up to $-1$~V on the Ho$^{\rm A}$ atoms [Fig.~\ref{manip}(a)] switches them to Ho$^{\rm B}$. As a result, in Fig.~\ref{manip}(b) all the atoms have the same apparent height and are adsorbed on the bridge site. We note that this operation is reversible, {\it i.e.}, positive bias ramping up to +1~V on top of a Ho$^{\rm B}$ transforms it back into Ho$^{\rm A}$. Conversely, further ramping with negative biases up to $-1.2$~V on top of Ho$^{\rm B}$ atoms irreversibly switch them to a new Ho$^{\rm C}$ species, see Fig.~\ref{manip}(c). With an apparent height of $141 \pm 3$~pm, this species appear much smaller than Ho$^{\rm A}$ or Ho$^{\rm B}$. Similar atomic manipulations yield to an equivalent sequence of adsorption sites also on two MgO layers. Using the unchanged atoms as reference, we identify the possible displacement of the switched adatoms by subtracting subsequent images. Figure~\ref{manip}(d), obtained by subtracting (a) from (b), shows asymmetric spots at the positions of the three circled atoms. The white arrows point into the direction of the displacement and indicate that the switched atoms have moved along two perpendicular directions, which correspond to the two possibilities of hopping from an O to bridge site. Interestingly, by subtracting Fig.~\ref{manip}(b) from (c), we observe that the switched atoms have been displaced perpendicularly to their previous direction, {\it i.e.}, from a bridge to an Mg site, see Fig.~\ref{manip}(e).\\

\begin{figure}[b!]
\includegraphics{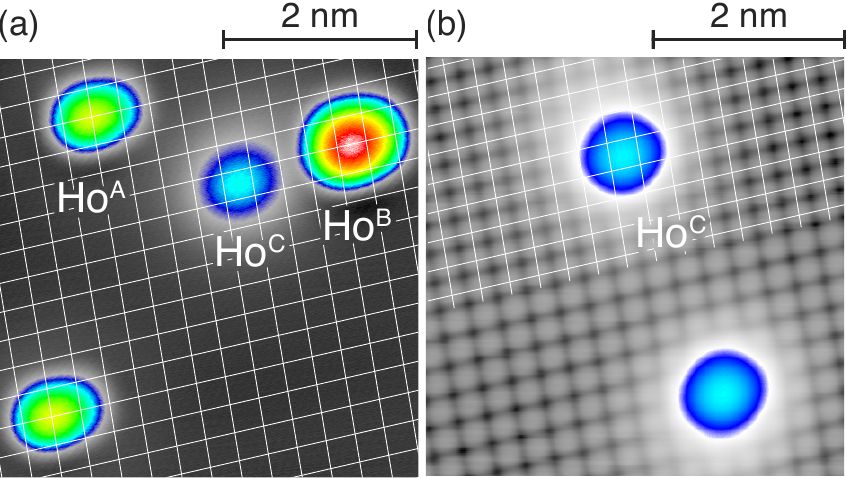}
\caption{(a) STM image of all three Ho species on 1 ML MgO ($V_\text{t} = -20$~mV, $I_\text{t} = 20$~pA). Ho$^\text{C}$ has an apparent height of $141 \pm 3$~pm and is adsorbed on Mg. The lattice is again inferred from an atomically resolved image of the bare MgO surface and marks the O atoms of the MgO. (b) Atomically-resolved image of two Ho$^\text{C}$ atoms on $2$~ML MgO.  The lattice marks the O atoms that appear as protrusions ($V_\text{t} = -20$~mV, $I_\text{t} = 5$~nA). Note that (b) has been recorded using the same tunneling parameters all along the image scan. The grid overlays the atomic protrusions, located at O lattice positions. Only half of the grid is shown to better distinguish the atomic protrusions in the lower part.}
\label{HoC}
\end{figure}

\begin{figure}[t!]
\includegraphics[width = 1.0\linewidth]{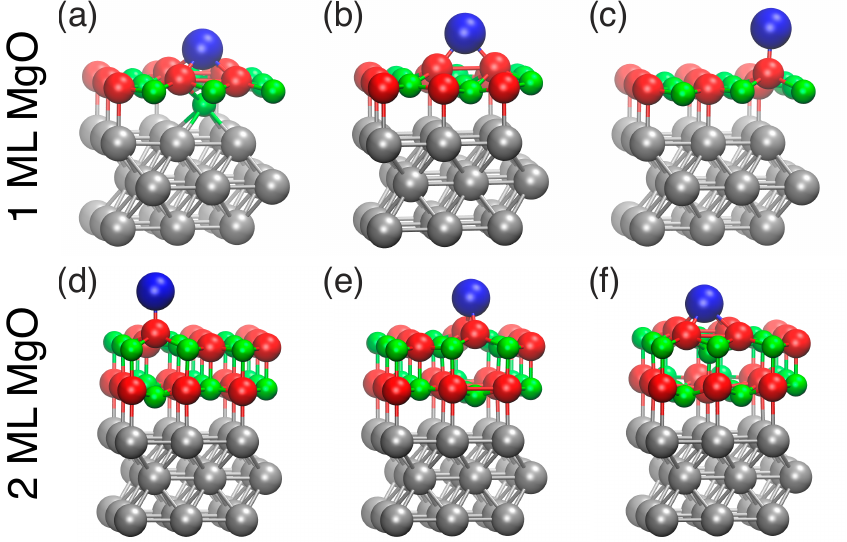}
\caption{DFT adsorption geometries of a Ho adatom on one (a-c) and two (d-f) MgO monolayer in order of decreasing binding energy from left to right. Color legend - red: O, green: Mg, gray: Ag, blue: Ho}\label{AdsGeometries} 
\end{figure}

Ho$^\text{C}$ adatoms are remarkably stable. Voltage ramps up to $\pm 10$~V have no effect on them, while Ho$^\text{A}$ or Ho$^\text{B}$ are transformed or desorbed under these conditions. Figure~\ref{HoC}(a) shows the adsorption site of Ho$^\text{C}$ to be on top of Mg. The grid marks the O sites and was determined with the same method as the one used for Fig.~\ref{sites}.\\
As shown by our DFT calculations for 1~ML MgO, this extraordinary stability results from a strong relaxation of the surrounding O neighbors and of the underlying Mg atom making this binding site $4$-fold O coordinated, see Fig.~\ref{AdsGeometries}(a). In agreement with experiment, this site has the highest binding energy, followed by the bridge and O sites, whose atomic geometries are shown in Figs. 7(b) and (c), respectively. See Table~\ref{dft} for the differences in binding energy and charge state of the atoms in the respective sites.

On two ML MgO, adsorption on top of Mg is calculated to be less favorable due to the presence of the subsurface MgO layer preventing large relaxation of the surface lattice, see Figure~\ref{AdsGeometries}(f). Nevertheless, experiment still finds this site as the most stable one, although only reachable after atomic manipulations. Note also that the order of the charge states is in agreement with the atomic manipulation from O via bridge to Mg sites requiring increasingly negative voltages. As also observed for Au and Ag on NaCl(100)\cite{rep04, ols07, ste15}, this indicates that the atoms become more and more positively charged along the transformation sequence.

The stability of the Ho$^\text{C}$ species enables imaging it under tunnel conditions that yield atomic resolution on MgO. These conditions displace or desorbs Ho atoms on the other two sites. The lattice overlaid onto Fig.~\ref{HoC}(b) marks the O atoms that appear as protrusions. According to our experience, this contrast is by far the most common one in low-bias images of the MgO surface. Only in very rare cases of tip chemistry and tunnel parameters are the Mg atoms imaged as protrusions. This clarifies the controversial DFT results on the STM contrast of MgO/Ag(100)~\cite{lop04, mal14, che14}.

\begin{table}[b]
\caption{DFT binding energy differences $\Delta E$ (eV) and charge transfers $\Delta q$ ($e$) for individual Ho, Au and Co atoms on O, bridge (br), and Mg sites on 1 and 2 ML thick MgO/Ag(100) and on the (100) surface of MgO bulk. The site with the highest binding energy is taken as reference; positive values signify a decrease in binding energy.}
\vspace{0.2 cm}
\begin{tabular}{|c|c|c|c|c|c|c|c|}
\hline
atom & site & \multicolumn{2}{|c|}{1 ML MgO/Ag} & \multicolumn{2}{|c|}{2 ML MgO/Ag} & \multicolumn{2}{|c|}{~~~~MgO(100)~~~~} \\\cline{3-8}
& & $\Delta E$ & $\Delta q$ & $\Delta E$ & $\Delta q$ & $\Delta E$ & $\Delta q$ \\
\hline
& O & 0.90 & 0.26 & 0.00 & 0.12 & 0.00 & $-$0.06 \\
Ho & br & 0.33 & 0.56 & 0.19 & 0.71 & 0.61 & $-$0.42 \\
& Mg & 0.00 & 1.20 & 0.21 & 1.21 & $-$ & $-$ \\
\hline
& O & 0.04 & $-$0.77 & 0.18 & $-$0.80 & 0.00 & $-$0.28 \\
Au & br & 0.00 & $-$0.77 & 0.00 & $-$0.83 & 0.15 & $-$0.31 \\
& Mg & 0.19 & $-$0.73 & 0.18 & $-$0.79 & 0.35 & $-$0.26 \\
\hline
& O & 0.00 & 0.14 & 0.00 & 0.01 & 0.00 & $-$0.13 \\
Co & br & 0.09 & 0.52 & 0.12 & 0.45 & 0.29 & $-$0.09 \\
& Mg & 0.52 & 0.92 & $-$ & $-$ & $-$ & $-$ \\
\hline
\end{tabular}
\label{dft}
\end{table}

\begin{figure*}[ht!]
\includegraphics{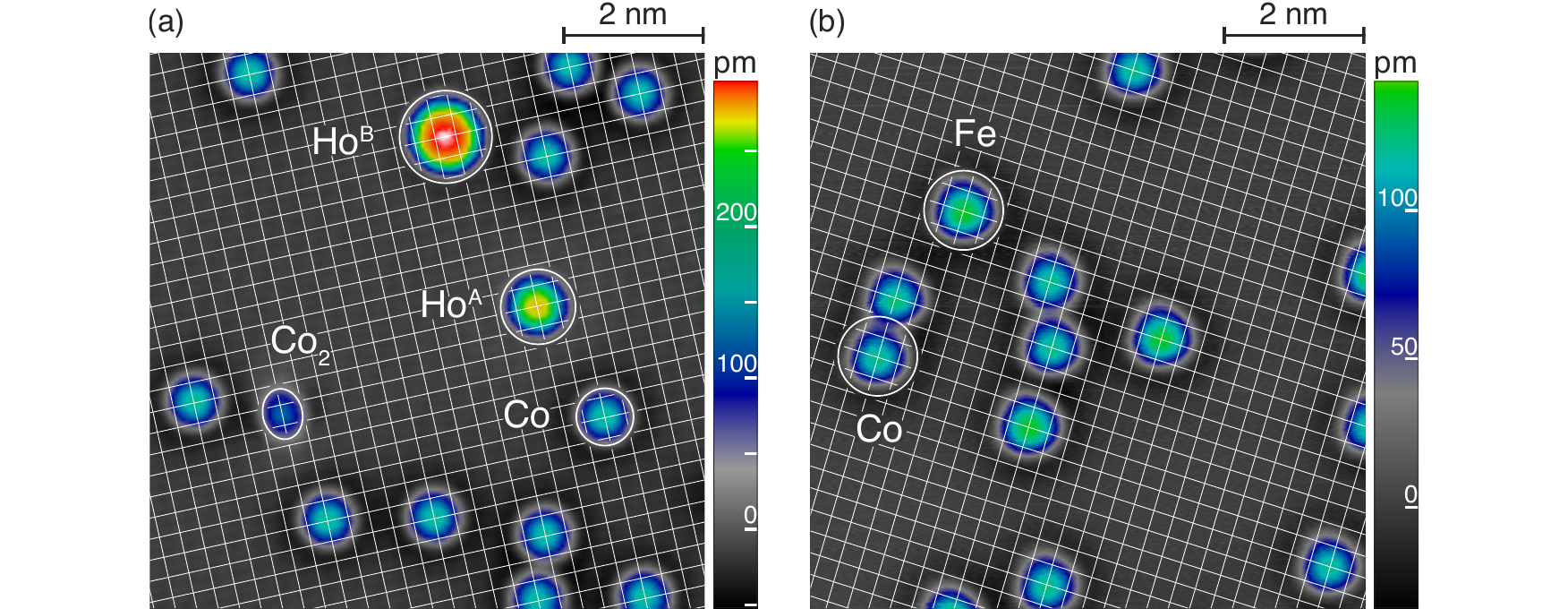}
\caption{STM images of (a) co-adsorbed Co and Ho and (b) Co and Fe on $1$~ML MgO(100)/Ag(100). The lattices are inferred from atomically resolved images of the bare MgO surface of the respective samples and mark the O positions. ((a) $V_\text{t} = -20$~mV, $I_\text{t} = 20$~pA, (b) $V_\text{t} = -50$~mV, $I_\text{t} = 20$~pA).}
\label{HoCoFeCo}
\end{figure*}
\subsection{Adsorption site of Co and Fe on MgO}
Our method of Ca doping can be applied to determine the adsorption site of any adatom on MgO. Once the site of one species is determined, one can use that species as marker to identify the sites of other atoms that are co-adsorbed. We use Ho atoms as marker for O and bridge sites to determine the adsorption site of co-adsorbed Co adatoms. Figure~\ref{HoCoFeCo}(a) shows a 1~ML thick MgO region with co-adsorbed Ho and Co atoms. The overlayed grid has again been extracted from atomically resolved images of the substrate and then been brought to coincidence with the Ho$^{\rm A}$ atoms. It therefore marks the O atoms and one sees that all Co atoms on that image are adsorbed on-top of O, in agreement with Ref.~\onlinecite{don16}.
Notice that Co is always on O, independent of the MgO layer thickness (up to $3$ layers) in excellent agreement with DFT. Using Co as a marker for the O sites, we further determine that Fe also adsorbs on top of O, Fig.~\ref{HoCoFeCo}(b), confirming former DFT calculations~\cite{mar03, ney04, che14, rau14, bau15b, alb15, don16}. Additionally, we note the presence of an elongated object of apparent height of $135 \pm 3$~pm, marked in Fig.~\ref{HoCoFeCo}. Scanning tunneling spectroscopy reveals inelastic steps at around 10~mV. Therefore, they are identified as Co dimers~\cite{bau15c}. We find that their axis is aligned along the O sublattice and each of the two constituent atoms is directly above or at least very close to an O site. This result indicates that our method can be extended to few-atoms clusters or small molecules.

\subsection{Adsorption sites of Au on MgO}
\begin{figure}[h]
\includegraphics{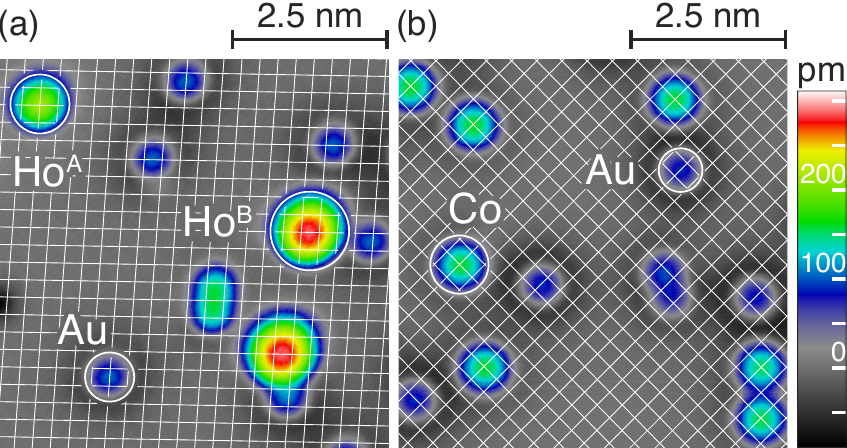}
\caption{STM image of co-adsorption of Au and Ho (a) and Au and Co (b) on $1$~ML MgO/Ag(100). The overlayed lattices mark oxygen positions in both cases. All Au atoms are found on the bridge site (a-b: $V_{\rm t} = -100$~mV, $I_{\rm t} = 20$~pA).}
\label{AuHo}
\end{figure}

We now apply our method to Au atoms, for which contradicting results were reported in the literature~\cite{pac05, ste07b, hon07}. Figure~\ref{AuHo}(a) shows 1 ML MgO with co-adsorbed Au and Ho, and a grid with lattice spacing and orientation being again extracted from an atomically resolved bare MgO spot of the same sample. Ho$^{\rm A}$ and Ho$^{\rm B}$ atoms are used to align the grid such that it represents the O positions. All the Au atoms shown in Fig.~\ref{AuHo}(a) are on bridge sites. A statistical analysis over 50 Au atoms on 1~ML MgO indicates that they almost exclusively adsorb on bridge sites, with a small fraction ($8 \pm 4$~\%) that is found on top of Mg. An equivalent identification done by co-depositing Co and Au atoms and using the Co atoms as reference for the O sub-lattice provides the same result within the error bars, see Fig.~\ref{AuHo}(b). Table~\ref{dft} shows that DFT finds the highest binding energy on the bridge site for mono- and bilayer MgO, and for the O site on MgO bulk, the latter supporting former EPR experiments~\cite{yul06} and DFT calculations~\cite{pac05}.

The use of adsorbates or substitutional atoms as atomic markers is the key for identifying an adatom's adsorption site on ionic or more general multi-element surfaces. STM images of Au atoms on 3 ML MgO were interpreted with an atomic lattice that was not calibrated with substitutional doping or other means, and a close to equivalent occupation of O and Mg sites was inferred~\cite{ste07b}. A small translation of this lattice by half of its lattice parameter would identify the two species as the two differently oriented bridge sites, thus compatible with our present finding and with former DFT calculations for Au atoms on 1-4 ML MgO/Mo(100)~\cite{hon07}. On mono- and bilayers MgO films, the presence of the substrate allows for an effective charge transfer to the adsorbed atom. Conversely, the charge transfer is reduced in absence of a metal support, as our DFT calculations and Refs.~\onlinecite{pac05, hon07} show. In addition to the reduced charge transfer to the substrate, DFT predicts a change of the adsorption site from bridge to oxygen going from ultrathin MgO films on Ag(100) to MgO bulk. This result is in agreement with former EPR experiment, which reported adsorption on top of O for 20~ML MgO/Ag(100)~\cite{yul06}. We therefore infer that the transition between the bridge and the O site occurs for MgO thickness above 3 ML~\cite{ste07b}.\color{black}

\section{Conclusion}
We presented a viable way to experimentally determine the adsorption site of adatoms on surfaces made of two or more elements and to interpret atomically resolved STM images thereof. For the specific case of MgO, we determine the adsorption sites of Ho, Co, Fe, and Au for the MgO mono- and bilayers grown on Ag(100). These results are of importance for the understanding of the fascinating electronic, catalytic, and magnetic properties of individual adatoms on thin films of ionic crystals.

\begin{acknowledgments}
We acknowledge support from the Swiss National Science Foundation under Projects 140479 and 157081, as well as the Serbian Ministry of Education and Science under grants ON171033 and ON171017. The DFT calculations were performed at the PARADOX-IV supercomputer at the Scientific Computing Laboratory (SCL) of the Institute of Physics Belgrade.
\end{acknowledgments}

\end{document}